%% file: main.tex
\documentclass[manuscript,nonacm]{acmart}

\AtBeginDocument{%
  }
\usepackage{makecell}
\usepackage{booktabs} 
\usepackage{amsmath}
\usepackage{graphicx}
\usepackage{multirow}
\usepackage[caption=false, font=footnotesize]{subfig}
\usepackage{xspace}

\DeclareRobustCommand{\company}{Company-X\xspace}
\DeclareRobustCommand{\system}{{\sc CMDBench}\xspace}

\newcommand{\stitle}[1]{\vspace{0.2em}\noindent\textbf{#1}}
\newcommand{\hide}[1]{}
\newcommand{\eg}{{\itshape e.g.}, }
\newcommand{\ie}{{\itshape i.e.}, }

\newcommand{\task}[1]{}

\begin{document}

\title{\system: A Benchmark for Coarse-to-fine Multimodal Data Discovery in Compound AI Systems}




\author{Yanlin Feng}
\authornote{Equal author contribution. Authors are ordered by random selection.}
\affiliation{%
  \institution{Megagon Labs}
  \country{USA}
  }
\email{yanlin@megagon.ai}

\author{Sajjadur Rahman}
\authornotemark[1]
\affiliation{%
  \institution{Megagon Labs}
  \country{USA}
  }
\email{sajjadur@megagon.ai}

\author{Aaron Feng}
\authornotemark[1]
\affiliation{%
  \institution{Megagon Labs}
  \country{Japan}
  }
\email{aaron@megagon.ai}

\author{Vincent Chen}
\authornotemark[1]
\affiliation{%
  \institution{Megagon Labs}
  \country{Japan}
  }
\email{vincent@megagon.ai}

\author{Eser Kandogan}
\affiliation{%
  \institution{Megagon Labs}
  \country{USA}
  }
\email{eser@megagon.ai}






\renewcommand{\shortauthors}{Feng et al.}

\begin{abstract}
  Compound AI systems (CASs) that employ LLMs as agents to accomplish knowledge-intensive tasks via interactions with tools and data retrievers have garnered significant interest within database and AI communities. While these systems have the potential to supplement typical analysis workflows of data analysts in enterprise data platforms, unfortunately, CASs are subject to the same data discovery challenges that analysts have encountered over the years --- silos of multimodal data sources, created across teams and departments within an organization, make it difficult to identify appropriate data sources for accomplishing the task at hand. Existing data discovery benchmarks do not model such multimodality and multiplicity of data sources. Moreover, benchmarks of CASs prioritize only evaluating end-to-end task performance. To catalyze research on evaluating the data discovery performance of multimodal data retrievers in CASs within a real-world setting, we propose \system, a benchmark modeling the complexity of enterprise data platforms. We adapt existing datasets and benchmarks in open-domain --- from question answering and complex reasoning tasks to natural language querying over structured data --- to evaluate coarse- and fine-grained data discovery and task execution performance. Our experiments reveal the impact of data retriever design on downstream task 
  performance --- $46\%$ drop in task accuracy on average ---
  across various modalities, data sources, and task difficulty. The results indicate the need to develop optimization strategies to identify appropriate LLM agents and retrievers for efficient execution of CASs over enterprise data.
\end{abstract}






\maketitle
\input{1_intro}
\input{2_overview}

\input{3_preparation}

\input{4_results}

\input{5_related}
\input{6_conclusion}


\bibliographystyle{ACM-Reference-Format}
\bibliography{paper}


\end{document}

%% file: 1_intro.tex
\section{Introduction}
\label{sec:intro}

The recent popularity of Large Language Models (LLMs) has led to the emergence of compound AI systems (CAS, hereafter)~\cite{compound-ai-blog} 
for accomplishing knowledge intensive tasks~\cite{petroni2021kilt,mishra2023characterizing}.
Examples of such systems include 
retrieval-augmented generation (RAG~\cite{lewis2020retrieval}) applications such as LlamaIndex~\cite{Liu_LlamaIndex_2022} and multi-agent platforms~\cite{kim2023llm, li2023camel, wu2023autogen}, where agents powered by LLMs interact with external tools and data retrievers to accomplish tasks. 
To this end, retrieval of data from domain-specific sources --- absent from an LLM's pre-training corpus such as those in enterprise data~\cite{gomez2017enterprise, armbrust2021lakehouse}  --- is a crucial task.
CASs consisting of proficient language agents such as LLMs and retriever agents can play a crucial role in the discovery and analysis of enterprise data by supplementing typical analysis workflows of data analysts~\cite{fernandez2018aurum,fernandez2016towards} or supporting ad-hoc end-user queries posed in natural language.
Enterprise data is often captured through multiple modalities such as tabular data with factoid information, graphs modeling symbolic knowledge, and documents with rich contextual information~\cite{10.1145/3211954.3211955}.
Therefore, the effectiveness of such compound AI systems --- such as LlamaIndex~\cite{Liu_LlamaIndex_2022} --- often relies on the 
efficiency of the retrievers that employ data discovery strategies to accomplish knowledge-intensive tasks such as search, QA, chat, fact-checking, and verification, among others. 

While compound AI systems necessitate instrumenting retrievers for multiple modalities, \ie tables, text, and graphs, data discovery beyond a single modality
is underexplored in existing literature. Existing work primarily considers unimodal data lakes such as tables~\cite{srinivas2023lakebench, fernandez2018aurum} and documents~\cite{Petroni2020KILTAB} on a variety of tasks~\cite{ Fan2022SemanticsawareDD,khatiwada2023santos, srinivas2023lakebench, saad2023ares, es2023ragas}. CMDL~\cite{eltabakh2023cross} and VerifAI~\cite{tang2023verifai} explore cross-modal data discovery over tables and short text documents. 
However, enterprise data may contain additional modalities such as domain-specific symbolic knowledge such as collection of triples that capture relationships/alignments among entities. 
Examples include domain-specific taxonomies, ontology, corporate directories, and social networks. The structured information encoded in such data has been shown to complement ML models in complex tasks~\cite{feng2020scalable,zhao2021knowledge, yasunaga2021qa}. 

\begin{figure}[!htb] 
  \centering
  \includegraphics[width=\linewidth]{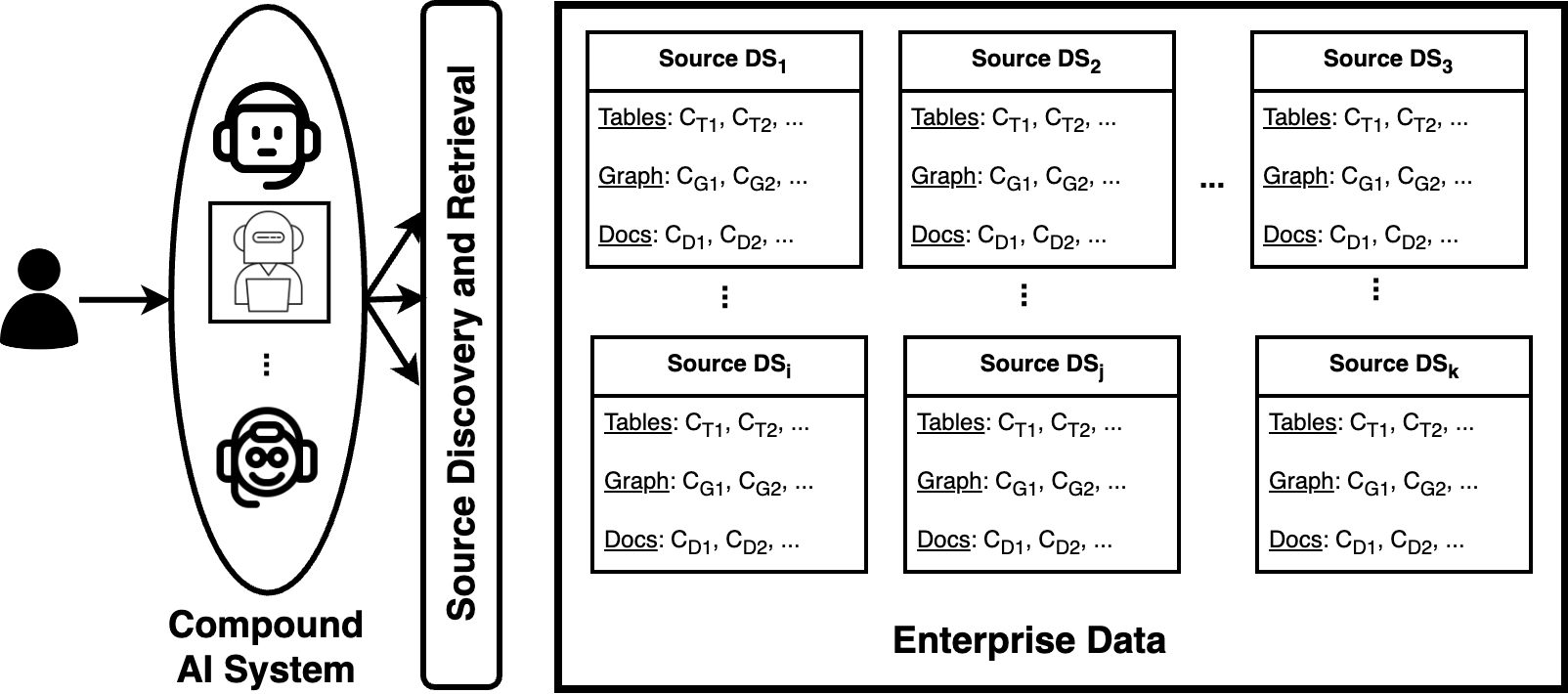}
  \caption{Each team-specific data source $DS_i$ represents a collection of sources ($S_m \in DS_i$) corresponding to different modalities ($m$) where $m \in \{D, G, T\}$ ($D$ = Document, $G$ = Graph, $T$ = Table). Collections of data $C_{mj} \in S_m$ may be stored in respective DBMSs --- tables in relational DBMS such as Postgres~\cite{stonebraker1986design}, text data in document DBMSs such as MongoDB~\cite{bradshaw2019mongodb}, graphs in property graph DBMSs such as Neo4j~\cite{miller2013graph}.}
  \label{fig:cas_enterprise} 
  \Description{MMD lake.}
\end{figure}

Moreover, the benchmarks discussed earlier are not accurate depictions of real-world enterprise data settings. Recent work~\cite{fernandez2018aurum} shows how modern organizations 
(\eg Merck, British Telecom, and the City of
New York) organize data into thousands of sources (\eg data lake, data warehouse, lakehouse~\cite{armbrust2021lakehouse}) that are managed by different
teams and departments and often become silos of information. Within such a setting, a CAS needs to address the same data discovery challenge~\cite{fernandez2016towards} as the analysts they are meant to complement --- not knowing which specific
\emph{data sources} (\eg $DS_1 - DS_k$ in Figure~\ref{fig:cas_enterprise})
contain the relevant data. Therefore, consideration of enterprise data impacts the choice of the benchmarks designed to evaluate multimodal data discovery performance within CASs~\cite{Liu_LlamaIndex_2022, compound-ai-blog}. 
We elaborate on the setting in Section~\ref{sec:overview}. Developing CASs --- capable of serving any user query over such enterprise data --- requires discovery of appropriate sources and modalities of information before performing retrieval. Existing unimodal and cross-modal (\ie text and tables) discovery methods employ custom-built or off-the-shelf vector indices that perform direct data retrieval~\cite{eltabakh2023cross,tang2023verifai}. As a result, a more in-depth coarse- and fine-grained evaluation of various discovery objectives, \ie source discovery and data discovery, is crucial for optimizing knowledge-intensive task execution in compound AI systems~\cite{compound-ai-blog}.


Constructing such multimodal data discovery benchmark is not straightforward as it requires creating a realistic representation of enterprise data sources, mapping and integrating multiple sources in a domain, employing strategies for ensuring appropriate coverage of all modalities, instrumenting quality control measures, designing benchmarking tasks and curating task instances, and preparing ground truth and provenance for performing a variety of task evaluations. To facilitate the evaluation of discovery performance in compound AI systems, where an agent may access specific information in one or more knowledge source modalities, we introduce \system --- a data collection and benchmark for multimodal data discovery. \system aims to lower the entry barrier for such research by formulating several data discovery tasks with respect to a common interface and a unified knowledge source, \ie Wikipedia. While LLM's are known to perform well in various tasks in the Wikipedia domain, we observed failure cases for tasks that require knowledge of disparate sources and advanced reasoning and computation. Consider the question ``\emph{Over the years, which team drafted the tallest players on average for the guard poisition?}'' We found all three popular LLMs GPT-4, Claude-3, and Gemini, fail to answer the question as it requires complex multi-hop reasoning over concepts such as players and teams coupled with aggregation and comparison tasks. Therefore, a more effective strategy is to employ a CAS with agents dedicated to specific tasks~\cite{kim2023llm}.


In this work, we report our current progress in developing such multimodal data discovery benchmark. 
In particular, we create a data source representing Figure~\ref{fig:cas_enterprise} in the NBA (national basketball association) domain adapting from and integrating existing sources of documents~\cite{Petroni2020KILTAB} and tables~\cite{zhongSeq2SQL2017} extracted from Wikipedia while introducing an additional modality, a knowledge graph extracted from Wikidata~\cite{vrandevcic2014wikidata}. We introduce discovery tasks for evaluation of unimodal discovery models and create task-specific natural language questions adapted from existing benchmarks of knowledge-intensive language tasks~\cite{Petroni2020KILTAB,zhongSeq2SQL2017,kwiatkowski2019natural,joshi2017triviaqa,cao2022kqa,yang2018hotpotqa}. We introduce an additional evaluation objective of source discovery and design benchmarks and baseline source discovery methods to report the impact of data discovery efficacy on the downstream task completion objective of a compound AI system. We also report preliminary results of our experiments with several discovery models adapted from the recently popularized \emph{LlamaIndex} framework~\cite{Liu_LlamaIndex_2022}. 
Our experiments highlight how data discovery performance varies with data modality, task complexity, and model design. In fact, we observed a $46\%$ drop in accuracy across all modalities even after employing the best-performing discovery models. The result indicates the emergence of a complex design space of data discovery agents within CAS, which requires continued benchmarking and monitoring while exploring trade-offs in dimensions such as accuracy, latency, and storage. We release \system at \url{https://github.com/megagonlabs/CMDBench}

%% file: 2_overview.tex
\section{Multimodal Data Discovery}
\label{sec:overview}


\subsection{Scope and Setting} 
Within an industry setting, project-specific teams working in an organization create custom sources of data, collected and often transformed through various workflows, over time. Such an observation is derived from the authors' experience working with enterprise data in the human resources (HR) domain at \company --- an industrial lab focusing on natural language processing, data management, and human-centered AI research. 
\company is a subsidiary of a large holding and conducts research and development for the other subsidiaries with worldwide businesses in staffing, HR, travel, marketing, and other online consumer services. Within the HR domain, a team working on matching candidates to companies may be interested in only resumes and job descriptions, structured data extracted from the text and their representations, company-specific statistics, and HR domain-specific knowledge, \ie relationships among different concepts such as jobs and benefits. Another team working on assisting job-seekers with search (\eg role and company) will not be interested in resumes, rather jobs and companies and search logs, among others. 

\begin{figure}[!htb] 
  \centering
  \includegraphics[width=\linewidth]{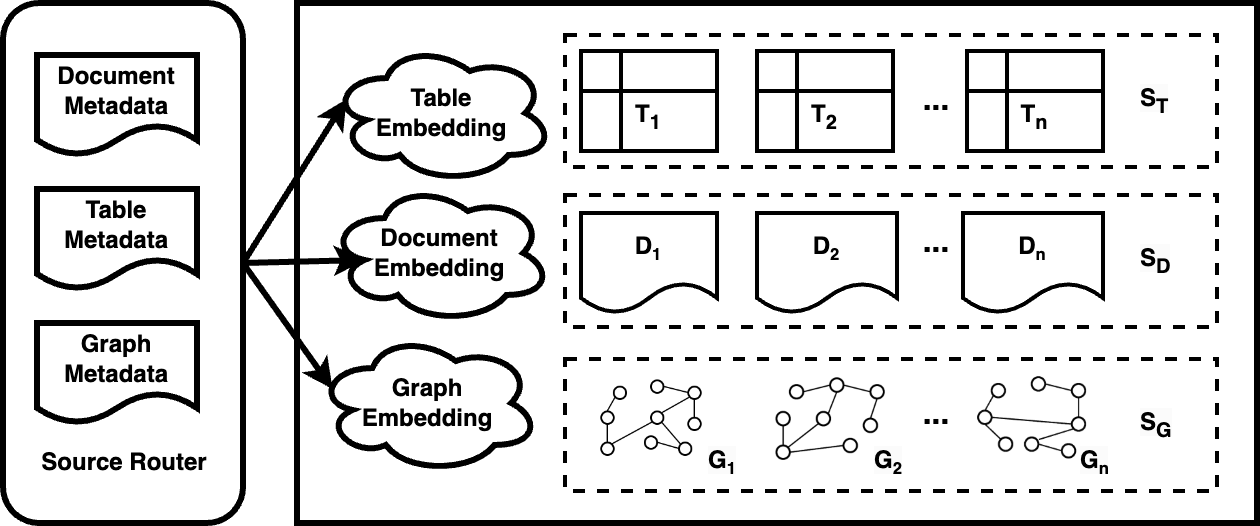}
  \caption{A simplified representation of the \system scope and setting with only one data source corresponding to basketball ($DS_b$). The tabular source ($S_T$) have statistics about players and teams, the graph source ($S_G$) contains symbolic knowledge and relationship among concepts, and the document source ($S_D$) has additional contextual information.}
  \label{fig:scope} 
  \vspace{-10pt}
  \Description{MMD lake.}
\end{figure}

Figure~\ref{fig:cas_enterprise} depicts the setting explored in this paper: \emph{compound AI systems operating over such enterprise data}. Each team-specific data source $DS_i$ represents a collection of sources ($S_m \in DS_i$) corresponding to different modalities ($m$)  --- such as multiple tables and documents --- created by specific teams. While enterprises may contain data corresponding to other modalities such as audio, video, and image, in this work we limit our scope to documents ($m=D$), tables ($m=T$), and graphs ($m=G$). Depending on the application and the platform architecture (\eg data warehouse or lakehouse~\cite{armbrust2021lakehouse}), the graph data may be stored as triples~\cite{bordes2013translating} to enable graph learning or in other format such as GraphAr~\cite{li2023enhancing}. However, to enable efficient querying and retrieval over different modalities within a compound AI system, collections of data ($C_{mj} \in S_m$) may be stored in respective DBMSs --- tables in relational DBMS such as Postgres~\cite{stonebraker1986design}, text data in document DBMSs such as MongoDB~\cite{bradshaw2019mongodb}, graphs in property graph DBMSs such as Neo4j~\cite{miller2013graph}. 

\stitle{Problem domain.} Let us consider the domain of all sports in United States such as basketball, football, and soccer, among others. Imagine a sport analytics company (\eg \emph{The Athletic\footnote{\url{https://theathletic.com/}}}) with different teams dedicated to modeling and analytics of each sport and creating their own data sources ($DS_i$). Within each team, there can be modality-specific sources $S_m$ containing multiple collections $C_{mj}$. For example, for basketball, there can be team-specific collections of tables with each collection stored in a dedicated database in Postgres. Similarly, team-specific documents and graphs may be stored in separate databases. Alternatively, the collections may be created based on men's (NBA) and women's (WNBA) basketball associations. In this work, we limit our scope to one sport, basketball (\ie data source $DS_{b}$). However, the setting, evaluation objectives, and benchmark construct discussed throughout the paper apply to multiple sources also. As shown in Figure~\ref{fig:scope}, there are three different information sources for basketball. While there might be information overlap among these sources, each offers unique information not contained in other collections. For example, the tabular source ($S_T$) has statistics about players and teams that the documents and graphs don't contain, the graph source ($S_G$) contains symbolic knowledge and relationships among concepts that tables and documents don't contain, and the document source ($S_D$) has additional contextual information and elaborations not included in either graph or tables. For simplicity, we assume the collections corresponding to each modality, \ie tables, documents, and graphs, are stored in modality-specific single databases. However, as mentioned earlier, these collections can be grouped to formulate additional collections on dimensions such as teams-specific collections and corresponding separate databases. Given the setting outlined above, we explain the data discovery tasks next.



\begin{figure*}[] 
  \centering
  \includegraphics[width=0.8\linewidth]{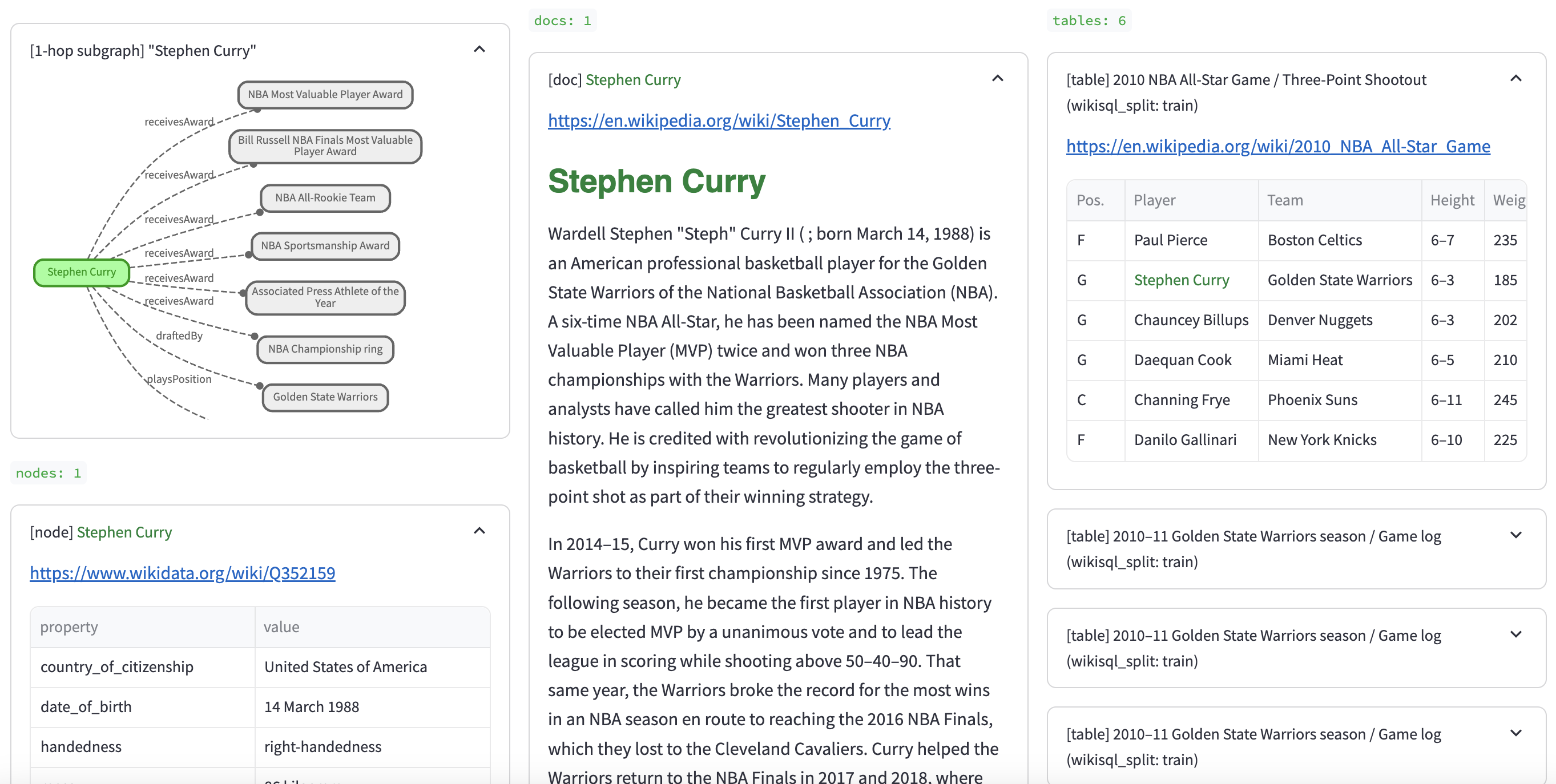}
  \caption{Snapshot of the NBA data source explored in \system showing 1-hop neighborhood of the \emph{player} Stephen Curry in Wikidata~\cite{vrandevcic2014wikidata}, his Wikipedia biography retrieved from the KILT document collection~\cite{petroni2021kilt}, and a table from WikiSQL~\cite{zhongSeq2SQL2017} with a list of NBA player appearances in different seasons. 
  }
  \label{fig:teaser} 
      \vspace{-10pt}
  \Description{MMD lake.}
\end{figure*}

\subsection{Data Discovery in Compound AI Systems} 
CASs such as LlamaIndex~\cite{Liu_LlamaIndex_2022} employ unimodal retrievers and query engines to retrieve data from a specific modality. 
However, various modalities that contain relevant information for a given task are not known beforehand. Consider the task of comparing the careers of Stephen Curry and Michael Jordan in terms of points scored, awards received, and contemporary opinions. Generating a response for this task, for example using a RAG application, requires discovering appropriate tables or sub-graphs, or documents not known beforehand to obtain the required information. Therefore, we define data discovery within enterprise data as a two steps process where high-level source discovery is performed first and then source-specific data discovery model(s) are employed for accomplishing a task. As shown in Figure~\ref{fig:scope}, a \emph{Source Router} leverages source metadata to route queries to one or more sources. Given a source, the appropriate collection can be retrieved using various approaches such as vector embedding-based search. Finally, collection-specific queries are issued to retrieve the relevant data. For example, the NBA data source snapshot shown in Figure~\ref{fig:teaser} displays the 1-hop neighborhood of Stephen Curry in the graph, his biography, and a table with a list of NBA player appearances in different seasons. DBCopilot~\cite{wang2023dbcopilot} explores a similar setting for NL2SQL tasks where answering a question via SQL requires identifying table(a) using an LLM-assisted schema-based router.


\stitle{Discoverable Elements.} The discoverable elements, \ie the abstract unit of
discovery~\cite{eltabakh2023cross, fernandez2018aurum}, can be both coarse-  and fine-grained. In this work, the coarse-grained element is a \emph{source} $S_m$. Within each source, examples of fine-grained elements include tables and columns for tabular data, documents and paragraphs for textual data, and nodes, edges, and sub-graphs for knowledge graphs. As mentioned earlier, to enable retrieval of the fine-grained discoverable elements we further store and index the data. We store tables in a Postgres database, and knowledge triples in a Neo4j graph database. Additionally, we maintain vector indices of text documents to support retrieval at various granularities. 


%% file: 3_preparation.tex
\section{Benchmark Preparation}
\label{sec:prep}


The primary goal of our work is to evaluate the multimodal discovery performance of retrievers utilized by agents in CASs. However, no common interface facilitates investigating discovery performance across sources and modalities. Moreover, within a single data source $DS_i$, the discovery tasks across sources $S_m$ are expected to be grounded on the same topic, \eg QA on NBA for $DS_b$ or QA on job recommendation for matching tasks in the HR domain. Consider the example of comparing the careers of Stephen Curry and Michael Jordan. In this case, while modality-specific retrievers may be employed, the retrieval goal is still limited to the domain of basketball. Therefore, a unified knowledge source capturing various modalities is required to characterize the multimodal discovery capabilities within a compound AI system. 
Wikipedia is an example of such a unified source. To build a common interface for evaluation, we leverage KILT~\cite{petroni2021kilt}, a knowledge source created to facilitate benchmarking a suite of knowledge-intensive tasks --- such as question answering, fact-checking, and entity extraction, among others --- on Wikipedia documents. To introduce other modalities, we integrate tables from the WikiSQL~\cite{zhongSeq2SQL2017} benchmark of NL2SQL tasks and triples extracted from the Wikidata~\cite{vrandevcic2014wikidata} graph.

\subsection{Data Collection} 
In this work, we limit our scope only to the basketball domain ($DS_b$)  instead of the entire Wikipedia for ease of data modeling, data preparation, and benchmark creation. As shown in Figure~\ref{fig:scope} and Table~\ref{tab:summary_tab}, even the domain of basketball provides a reasonable diversity in data in different modalities that can support a variety of queries over CASs from factoid and aggregate questions to knowledge-intensive tasks requiring multi-hop reasoning over one or more sources of information. 

\input{tables/statistics}

\subsubsection{Text Data}
To collect text documents on the topic of NBA, we select the KILT~\cite{petroni2021kilt} collection of approximately six million Wikipedia pages. We selected KILT due to its rigorous standardization and quality control measures to create the collection --- from using a consistent Wikipedia snapshot to preserving the provenance of answers to questions from a suite of Wikipedia-based benchmarks on a variety of tasks such as question answering, fact-checking, and dialogue system. The collection also contains additional metadata of these pages, such as titles and categories. We  select 8314 pages related to the NBA category. 


\subsubsection{Tabular Data} 
The KILT collection, however, does not contain any tabular data --- the tables in Wikipedia pages are removed during the standardization process. To ensure that all modalities capture the same underlying concept of data source from Wikipedia, \ie basketball, we select the WikiSQL benchmark~\cite{zhongSeq2SQL2017} of NL2SQL tasks. The WikiSQL benchmark was constructed by sanitizing the HTML tables in the WikiTables~\cite{bhagavatula2013methods} collection and then adding approximately 80,000 questions-SQL pairs created via a combination of automated generation and crowdworker validation. 
Each table also contains as metadata the title of the Wikipedia page it was extracted from. In this work, we select a subset of these tables (1076 tables) whose page titles appear in the NBA document collection created above. Besides the page title (\eg 2021–22 Los Angeles Lakers season), each table optionally contains other metadata such as section title (Player stats) and caption (\eg Regular season statistics.) 

 \begin{figure}[!htb] 
  \centering
  \includegraphics[width=0.5\linewidth]{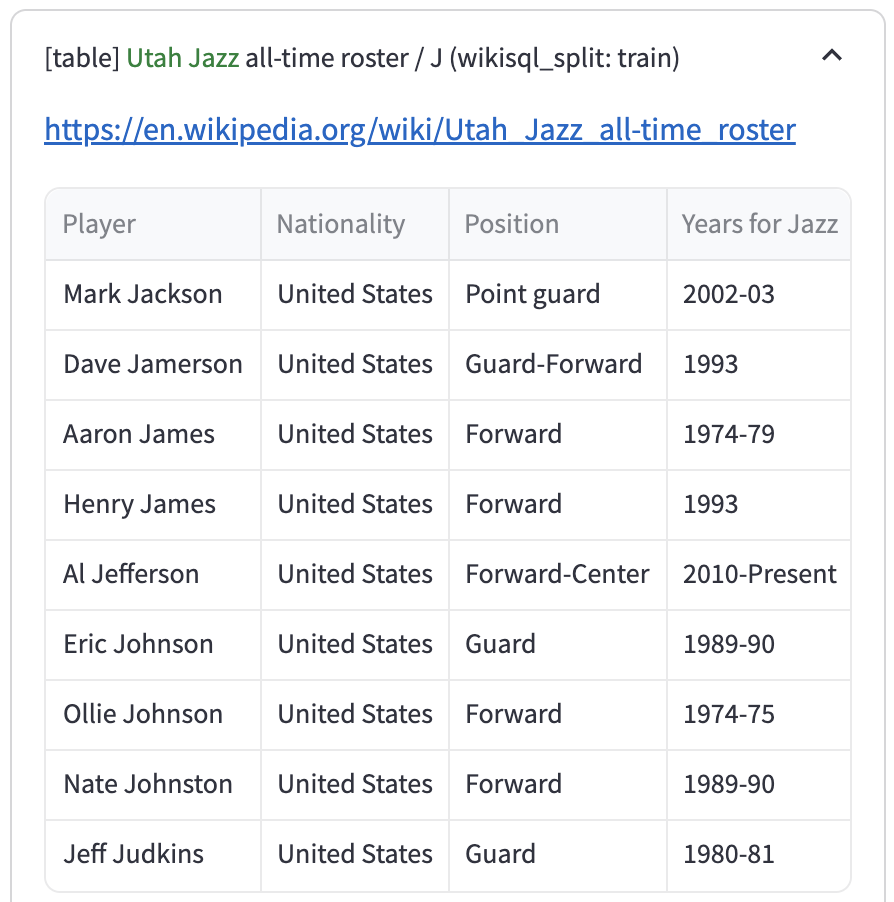}
  \caption{WikiSQL table corresponding to $Q3$ in Table~\ref{tab:tabular-task}.}
  \label{fig:table_nba} 
  \Description{MMD lake.}
\end{figure}

Upon empirical observation, we identified that some of the tables contained incorrect captions and section titles. Therefore, we parsed the corresponding Wikipedia pages to update the metadata. 
In addition to the raw tables, SQL queries, results, and natural language questions, 
The WikiSQL benchmark also released a corresponding SQL database. However, in the original release the database column names of tables were symbolized, \ie original column names in Wikipedia tables were renamed to identifiers such as $col0$ and $col1$. Instead of using the default database, we created a new database from the source JSON files in the benchmark by transforming each JSON document to a valid PostgreSQL table observing the original schema and column type constraints. Figure~\ref{fig:table_nba} shows a table in the NBA collection of \system.



 \begin{figure}[!htb] 
 \vspace{-10pt}
  \centering
  \includegraphics[width=0.5\linewidth]{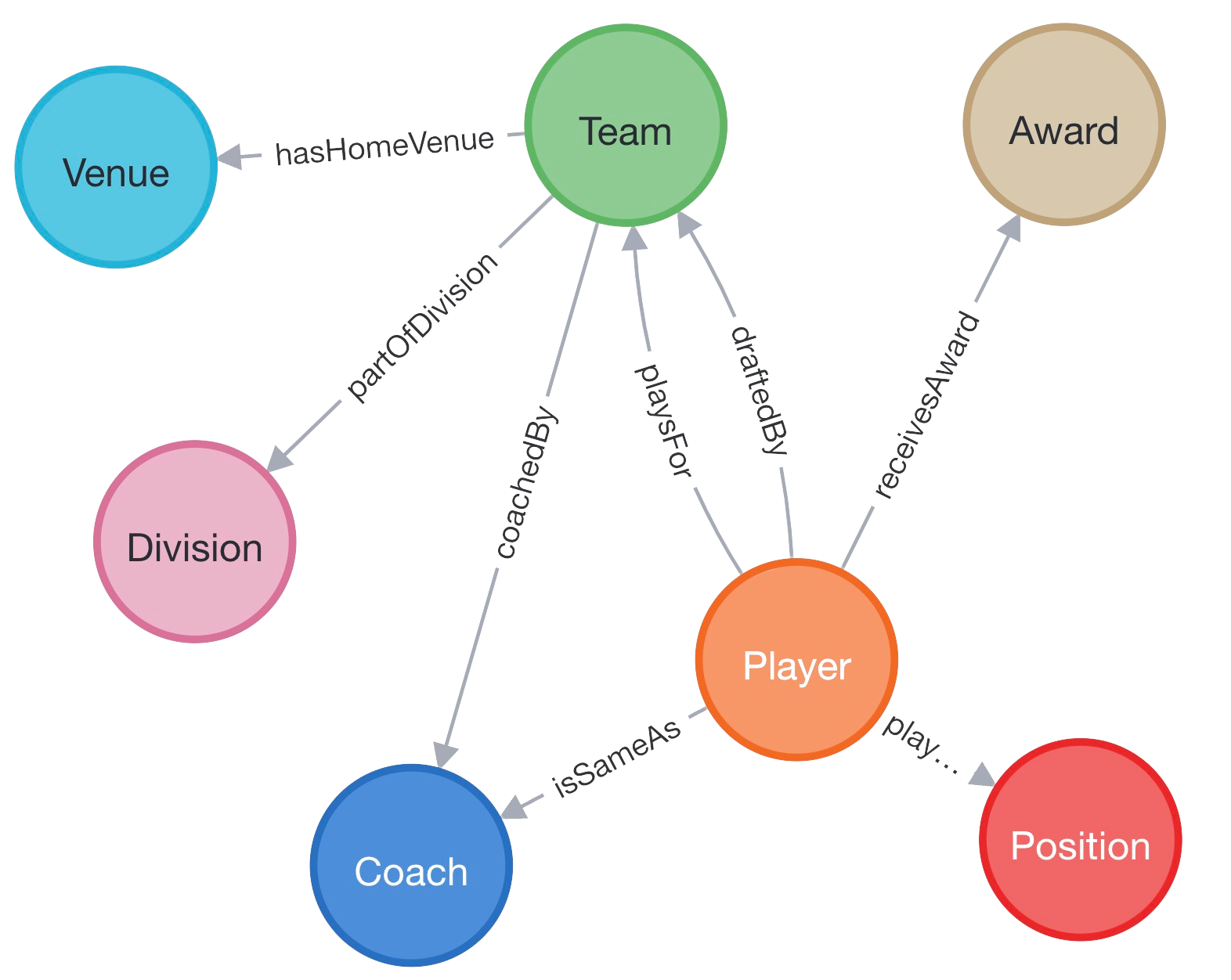}
  \caption{Schema of the NBA graph extracted from Wikidata.} 
  \label{fig:graph_nba} 
  \Description{MMD lake.}
\end{figure}

\subsubsection{Graph Data} 

To construct a knowledge graph in the NBA domain, we again referred to the KILT collection, which uses T-Rex~\cite{elsahar2018t}, a large-scale collection of facts (\ie knowledge triples) aligned to sentences in Wikipedia. However, our initial experiments revealed very low coverage of triples in the NBA domain, specifically due to the fact that T-Rex was created using Wikipedia abstracts only. Therefore, we created the NBA graph from scratch via extraction from Wikidata~\cite{vrandevcic2014wikidata} --- an open-source large-scale knowledge graph. We designed the schema of the knowledge graph
(see Figure \ref{fig:graph_nba}) 
with a specific focus on capturing concepts related to sports teams.
Due to our choice of Neo4j as the graph DBMS, which employs a property graph model, we also hand-picked several properties for each node and relation type, \eg height and weight of players. We extracted a sub-graph by running SPARQL queries on the Wikidata SPARQL endpoint. Next, we extracted the relations and properties by running a separate SPARQL query for each relation or entity property. Finally, we ingested the resulting graph into a Neo4j by transforming triples to a property graph.


\input{tables/task_example}

\subsection{Task Design}
The goal of the benchmark is to evaluate coarse- and fine-grained discovery performance of respective models/agents. Therefore, we construct goal-specific benchmarks via a combination of automatic selection of tasks (\ie questions) from existing benchmarks and manual annotation of various aspects such as task types and provenance of retrieved information at various granularities of discoverable elements --- coarse-grained source $S_m$ and fine-grained elements (\eg table and column, document and paragraph, and node and path.) 
However, to understand the impact of data discovery efficacy on the downstream task, it is important to evaluate the end-to-end task performance. Therefore, we explore the three high-level goals coarse-grained source discovery ($DE_c$), fine-grained data discovery ($DE_f$), and task execution accuracy.

\stitle{Source Discovery.} Evaluate whether, given a task, the appropriate coarse-grained discoverable element, \ie source(s), can be discovered. In this benchmark, there are three sources: Wikipedia documents, WikiSQL tables, and the WikiData graph. In Table~\ref{tab:tabular-task}, we show examples of questions and the corresponding sources. Note that a question may be answered using elements of more than one source, \eg the final question in Table~\ref{tab:tabular-task} ($Q7$.)

\stitle{Data Discovery.} Given the source, evaluate whether all the relevant fine-grained elements can be discovered for the given task. For example, answering $Q1$ and $Q2$ in Table~\ref{tab:tabular-task} requires discovering one and two documents, respectively. While not shown in Table~\ref{tab:tabular-task}, we also evaluate \emph{paragraph discovery} performance, \ie given a question and the document, whether the paragraph containing the answer can be discovered. Similarly, fine-grained discovery tasks on graph modality are identifying the node ($Q4$), edge ($Q5$), and sub-graph ($Q6$) containing the answer to a question. For tabular data the fine-grained elements are table and column ($Q3$.)



\subsection{Workload Creation}
We now explain how the benchmark is created by selecting questions from existing QA datasets~\cite{Petroni2020KILTAB,zhongSeq2SQL2017,kwiatkowski2019natural,joshi2017triviaqa,cao2022kqa,yang2018hotpotqa}. For each question, two annotators (authors of the paper) independently verified the ground truth source --- both coarse- and fine-grained --- and the degree of complexity of the question. The annotators then resolved any disagreements to create the eventual benchmark.

\subsubsection{Document Discovery} 
We selected questions from three open domain question-answering datasets where the answers are short, concise phrases --- TriviaQA~\cite{joshi2017triviaqa}, Natural Questions~\cite{kwiatkowski2019natural}, and HotpotQA~\cite{yang2018hotpotqa} --- used in the KILT benchmark~\cite{petroni2021kilt}. 
In HotpotQA, each question is associated with two paragraphs and requires multi-hop reasoning to infer the answer, while for TriviaQA and Natural Questions, the question is answerable from a single document (paragraph.) For each question, KILT includes the provenance of the source document(s) required to provide the answer. We leveraged the provenance information to first identify 120 questions related to the NBA domain. After manual verification of source provenance, we excluded questions that were ambiguous or time-sensitive --- (\eg ``Who won the NBA scoring title this year?'') --- resulting in 63 questions for document discovery. However, not all of these questions are exclusively answerable by documents. We manually annotated additional sources and found that $25\%$ of the questions were answerable by more than one source.




\subsubsection{Table Discovery}
To create the questions for table discovery, we first examined the questions in WikiSQL~\cite{zhongSeq2SQL2017} and found that the majority of the questions were ambiguous and difficult to understand without prior knowledge of the corresponding table in the ground truth, \eg When was the score 123-112?
Therefore, we aim to augment the questions with contextual information to aid in discovering tables and columns (see Table \ref{tab:tabular-task} for examples.)
In this paper, we only select questions that can be answered by a single table.
Given a collection of questions in WikiSQL~\cite{zhongSeq2SQL2017} and the corresponding ground truth tables, we employ a retrieval-augmented question generation pipeline --- leveraging GPT-3.5 Turbo --- to generate 100 questions.
We manually examined each question and selected 84 questions that were deemed to be valid, unambiguous, and answerable questions for table discovery by two independent annotators.
For each question, we crafted a ground truth SQL, and added the corresponding table and columns as ground truth discoverable elements in the provenance set. 
Note that the questions can be further grouped into two categories: easy (\ie factoid questions --- $Q3$) and hard (\ie aggregate questions with additional operations such as sort and filter --- $Q4$). Out of the 67 questions, $8\%$ were answerable by more than one modality.

\subsubsection{Graph Discovery}
For graph discovery evaluation, we selected questions from the KQA pro dataset~\cite{cao2020kqa}, a knowledge base question answering dataset with WikiData as its knowledge source. From the questions that can be answered using our NBA knowledge graph, we excluded ambiguous (\eg ``Is the inception time not in 1895 for the basketball team that has member Jack Thompson (whose height is 185 centimeters)?'') or time-sensitive questions. We manually annotated each of the selected 36 questions to the add ground-truth discoverable element in the provenance set and handcrafted the corresponding gold Cypher query. Out of the 36 questions, $58\%$ were answerable by more than one modality.


\subsection{Evaluation Metrics}
As mentioned earlier, we aim to evaluate (1) performance in discovering relevant elements and (2) task execution effectiveness.


\stitle{Task performance.}
For all tasks, we use \emph{Accuracy}.
Answers questions over tables and graphs result in a discrete output, \ie result of executing natural language to domain-specific query translations over the respective modalities. For document questions answering, answers are extractive~\cite{joshi2017triviaqa,kwiatkowski2019natural} or short
abstractive~\cite{yang2018hotpotqa}. 

\stitle{Discovery performance.}
For various data discovery tasks within a source, we measure $R$-precision, calculated as $\frac{r}{R}$, where $R$ is the
number of discoverable elements in the provenance set of a given task and $r$ is the number of relevant elements
among the top-$R$ retrieved elements. Note that $R$-precision = Precision@1 when $R = 1$. We also measure Precision@$1$, Recall@$k$, where $k \in \{1,3,5\}$ for different sources and corresponding fine-grained discoverable elements. For both metrics, we report the mean over all tasks. We employ standard multi-label classification metrics to measure source discovery performance where each source type --- document, table, and graph --- corresponds to one class. For example, ``Graph\_P'' means, for questions that a model predicts ``graph'' as the source, how many of these are answerable using the graph? ``Graph\_R'' means, for answerable questions using the graph, how many of these did a model predict ``graph'' as the source?


%% file: tables/statistics.tex
\begin{table}[]\scriptsize
    \centering
    \caption{Summary of data in different sources.} 
    \vspace{-5pt}
    \begin{tabular}{c p{6cm}}
    \toprule
       \textbf{Source}  & \textbf{Statistics} \\ \midrule
       Documents~\cite{petroni2021kilt}  & 8314 documents on player profiles, teams including all start teams, all NBA seasons, and all time roasters, among others. \\
       Tables~\cite{zhongSeq2SQL2017}  & 1076 tables on different aspects such as roasters (163), game logs (544), player stats (19.) \\
       Graph~\cite{vrandevcic2014wikidata}  & 6265 nodes and 21556 relations with 4959 players, 38 teams, 1002 coaches, and 125 venues. \\
\bottomrule
    \end{tabular}
    \label{tab:summary_tab}
    \vspace{-10pt}
\end{table}

%% file: tables/task_example.tex
\begin{table*}[!htb]
\scriptsize
\centering
\caption{Example of data discovery tasks.}
\vspace{-5pt}
\begin{tabular}{p{4cm} p{5cm} c p{5cm}}
\toprule
Question & $DE_f$ & $DE_c$ & Example Query \\
\midrule
$Q1.$ What is Magic Johnson's real first name? & [``Earvin "Magic" Johnson Jr. (born August 14, 1959) is an American retired professional basketball player...''] & Document &
\texttt{search(``Magic Johnson'')} \\
$Q2.$ When was the 1993-94 Washington Bullets head coach born? & [``The 1993-94 NBA season was the Bullets' 33rd season in the National Basketball Association. ... and head coach Wes Unseld was fired.'', \newline ``Westley Sissel Unseld (born March 14, 1946) is an American former basketball player...''] & Document &
\texttt{search(``1993-94 Washington Bullets season''), search(``Wes Unseld'')} \\
\midrule
$Q3.$ Name a player with last name starting with J, who has played for the Utah Jazz for more than 4 years & table name = \texttt{utah\_jazz\_roster\_J}, \newline column name = \texttt{\{player, yrs\}} & Table &
\texttt{SELECT player \newline
FROM utah\_jazz\_roster\_J  WHERE yrs > 4 \newline} \\
$Q4.$ Who had the top-3 assists on the Lakers in season 2021-22? & table name = \texttt{player\_stats\_lakers\_2021\_2022}, \newline column name = \texttt{\{player, assists\}} & Table &
\texttt{SELECT player \newline
FROM player\_stats\_lakers\_2021\_2022 \newline
ORDER BY assists DESC LIMIT 3} \\
\midrule
$Q5.$ What is the height of LeBron James? & \raisebox{-0.4\height}{\includegraphics[height=0.6cm]{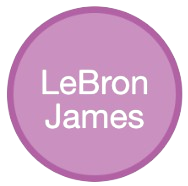}} & Graph & \texttt{MATCH (p:Player \{name: `LeBron James'\}) \newline RETURN p.height} \\
$Q6.$ When did Kevin Garnett join the Minnesota Timberwolves? &  \raisebox{-0.6\height}{\includegraphics[height=0.6cm]{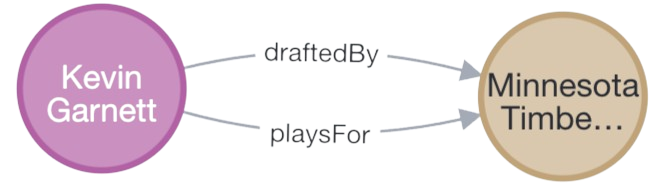}} & Graph & \texttt{MATCH (p:Player \{name: `Kevin Garnett'\})- \newline [r:playsFor]->(t:Team \{name: `Minnesota  \newline Timberwolves'\})  RETURN r.start\_time} \\
$Q7.$ In 1996, did any member of Orlando Magic weigh less than 150 kg? & \raisebox{-0.8\height}{\includegraphics[height=1.2cm]{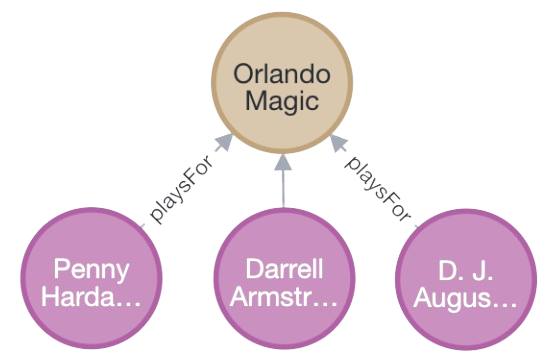}} & Graph & \texttt{MATCH (p:Player)-[r:playsFor]->(t:Team) WHERE t.name = `Orlando Magic' AND 2r.start\_time <= `1996' AND (r.end\_time >= `1996' OR r.end\_time IS NULL) ...} \\
\midrule
$Q8.$ What seasons that Justin Harper played for Orlando Magic? & [``...On December 9, 2011, he signed with the Magic. On October 27, 2012, he was waived by the Magic.''], \newline \newline table name = \texttt{justin\_harper\_nba\_regular\_season}, \newline column name = \texttt{\{year, team\}}, \newline \newline \includegraphics[height=0.6cm]{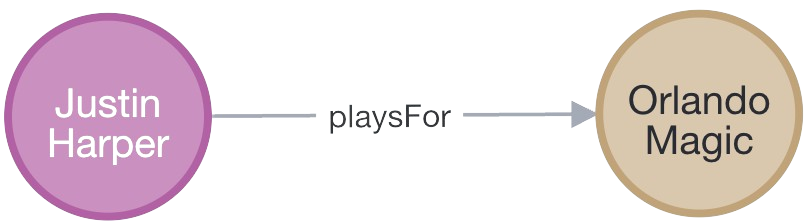} & All &
\texttt{search(``Orlando Magic Justin Harper'')}  \newline \newline
\texttt{SELECT year \newline
FROM justin\_harper\_nba\_regular\_season \newline
WHERE team = 'Orlando'}  \newline \newline
\texttt{MATCH (p:Player \{name: `Justin Harper'\})-  \newline [r:playsFor]-(t:Team \{name: `Orlando  \newline Magic'\}) RETURN r.start\_time, r.end\_time} \\
\bottomrule
\end{tabular}
\label{tab:tabular-task}
\end{table*}

%% file: 4_results.tex
\section{Experiment Results}
\label{sec:exp}
In this paper, we employed LlamaIndex~\cite{Liu_LlamaIndex_2022} as the compound AI system --- experimenting with available data discovery agents and implementing new baselines as needed --- to analyze coarse- and fine-grained data discovery performance on our designed benchmark. We selected LlamaIndex due to its popularity ($30.2$k GitHub stars and $4k$ forks), active maintenance status ($840$ contributors and $338$ releases), and support for wide-ranging agentic workflows.


\stitle{Baseline: Data Discovery.} We used three different embedding models to build document vector indices --- \emph{BM25}~\cite{robertson2009probabilistic}, BGE (base and large)~\cite{luo2024bge}, Ada-002 (the default GPT-based embedding in LlamaIndex.) Each vector index serves as one document retrieval baseline. We use the same vector indices for tables, except BGE-large. However, besides each table's title and column headers, we also use additional metadata, such as the corresponding Wikipedia page title, section title, and table caption, to provide more contextual information for embedding computation. For column discovery, we generate the NL2SQL query given the question and the table and extract appropriate column values for comparison. For graphs, we employ NL2Cypher queries to discover appropriate elements, similar to tables. We use two variations of answer generation LLMs for all data discovery baselines: GPT-3.5 Turbo and GPT-4 Turbo. 

\input{tables/source_selection}

 \begin{figure}[!htb] 
  \centering
  \includegraphics[width=0.7\linewidth]{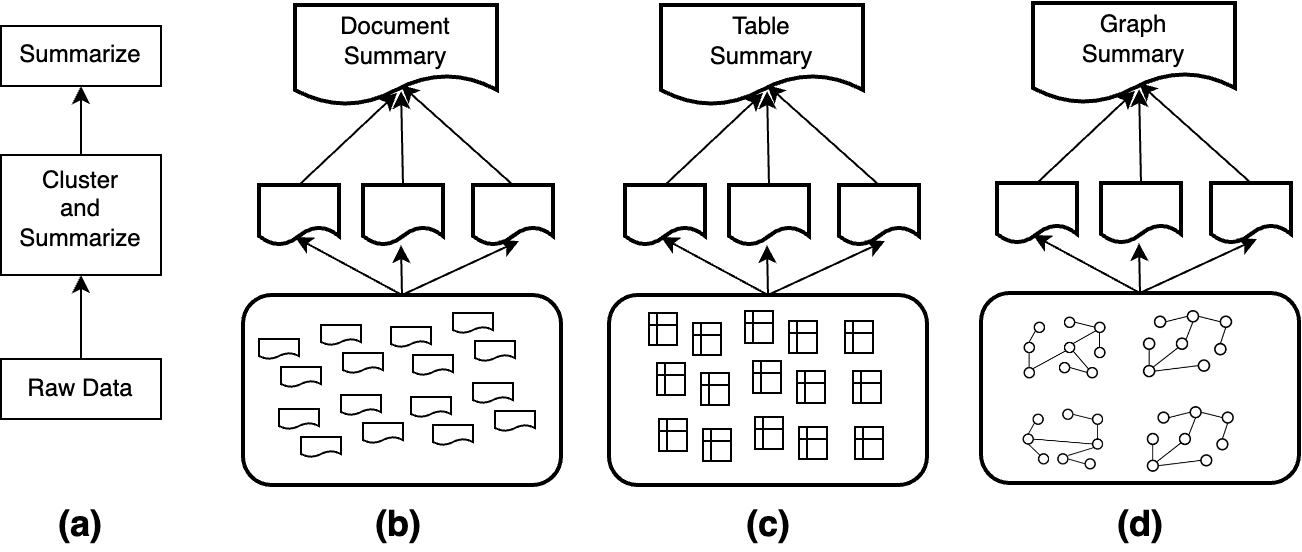}
  \caption{Overview of Source Discovery baseline workflow.}
  \label{fig:raptor} 
  \Description{MMD lake.}
\end{figure}

\stitle{Baseline: Source Discovery.} Our design of the source discovery baseline is inspired by the \emph{RouterQueryEngine} in LlamaIndex, which leverages source descriptions to route queries to relevant sources. However, without high-quality source descriptions, the source discovery model may incorrectly route queries to irrelevant sources. To address this issue, we adapt the idea proposed by RAPTOR~\cite{sarthi2023raptor}, which employs recursive embedding, clustering, and summarizing chunks of text to construct a tree with differing summarization levels from the bottom up. 
As shown in Figure~\ref{fig:raptor}, 
We employ a two-step process first to prepare cluster summaries and then perform another summarization to create the source description. For each modality, we create two different variations of source summaries --- basic ($sum_{b}$) and enhanced ($sum_{e}$) --- by varying the content we include in summary computation. For documents, we employ \emph{page titles} and \emph{page title + page abstract} to compute basic and enhanced summaries, respectively. For tables, we employ \emph{title} and \emph{title + column headers} to compute basic and enhanced summaries, respectively. Finally, for graphs, we verbalize the \emph{schema} consisting of nodes and relations only for basic summaries and additionally include node and relation property names to compute enhanced summaries. We employ several source selection strategies: (a) \emph{Select-$m$}: always select a specific source modality irrespective of the question, (b) \emph{Select-random}: randomly select a source modality, (c) \emph{Select-all}: select all modalities irrespective of the questions, (d) \emph{LLM}+$\mathbf{sum_{b}}$: given a question, an LLM (\eg GPT-3.5-turbo or GPT-4-turbo) selects one or more sources based on basic source summaries, and (e) \emph{LLM}+$\mathbf{sum_{e}}$: given a question, an LLM (\eg GPT-3.5-turbo or GPT-4-turbo) selects one or more sources based on enhanced source summaries.


\subsection{Results and Takeaways}

\subsubsection{Coarse-grained Source Discovery}
Table~\ref{tab:source_selection} provides an overview of the source discovery performance of various models under different summarization strategies. Unsurprisingly, LLMs outperform other simpler selection strategies. Overall, there's no significant difference in source discovery performance between GPT-3.5-turbo and GPT-4-turbo. For both LLMs enhanced summaries yielded better discoverability. However, our fine-grained analysis reveals that the enhanced summarization approach is only effective for documents and did not significantly improve the discoverability of graphs and tables over basic summaries. In fact, for graphs, the performance degraded with basic summaries. Since the detailed summary of the graphs includes information about node and relations properties such as height and weight, any questions on such properties are routed to the source by the LLM. However, for many lesser-known players, such information is missing in the graph, leading to lower precision. On the other hand, a detailed summary of documents contained rich contextual information from the abstracts, leading to higher-quality summaries. 

\begin{figure}[!htb]
    \centering
    \includegraphics[width=\linewidth]{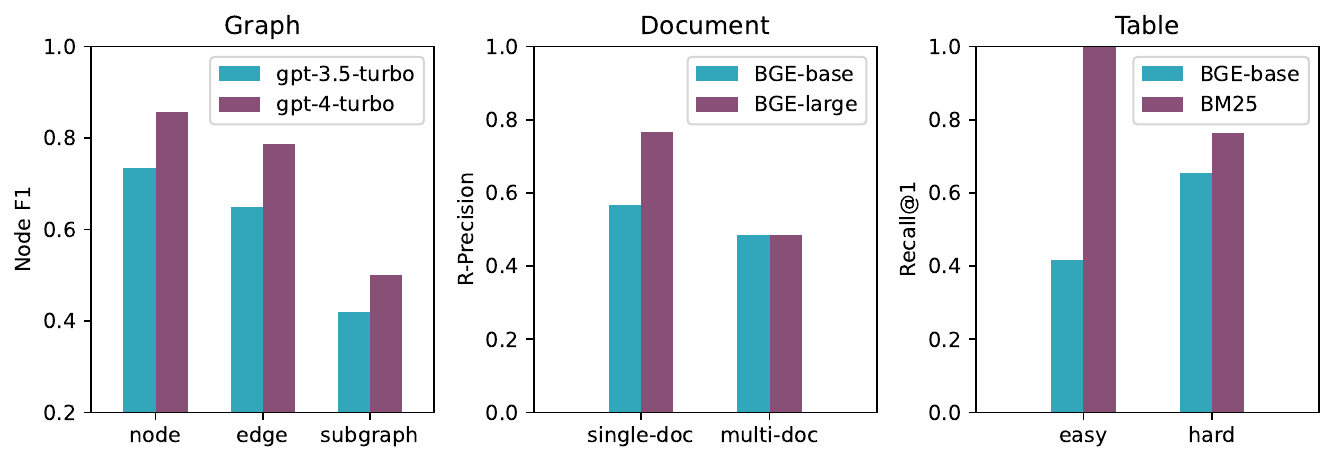}
    \caption{Source discovery on various question categories. }
        \vspace{-10pt}
    \label{fig:per_category}
    \vspace{-10pt}
\end{figure}

Figure~\ref{fig:per_category} captures the impact of question difficulty on the source discovery performance for various modalities --- except for BGE-base embedding for Tables, the performance degrades with increasing question difficulty across all modalities. For tables, difficult questions tend to contain more contextual information, such as column headers, which may result in better table retrieval than easy questions, which tend to be terse. We show $R$-precision for documents due to the presence of multi-document HotpotQA~\cite{yang2018hotpotqa} questions. We report Node-$F1$ for graphs as we aim to measure the overlap between the retrieved sub-graph and ground truth. Since all of our tabular discovery tasks involved retrieving only one table, we report Recall$@1$.

\subsubsection{Fine-grained Data Discovery}


Tables~\ref{tab:table_emb},~\ref{tab:fine_nl2cypher}, and~\ref{tab:doc_para} capture the performance of various embedding models for fine-grained discovery over tables, graphs, and documents, respectively. For document and paragraph discovery (Tables~\ref{tab:doc_para}), sparse retrieval methods such as BM25 perform poorly compared to dense retrieval methods such as BGE. Moreover, the larger the embedding models, the better performance. Somewhat surprisingly, BM25 achieved good performance in table discovery. Upon further reflection, we posit that embedding computed on table titles and captions, which are more keyword-heavy, lack rich contextual information for effective dense retrieval. On the other hand, such keyword-heavy content is more suitable for retrieval using the BM25 method. For graphs, we additionally explored the open LLMs such as Llama and observed poor NL2Cypher generation performance for edge discovery tasks.

\input{tables/table_discovery}
\input{tables/nl2cypher}
\input{tables/doc_para_discovery}





\begin{figure}[!htb]
    \centering
    \includegraphics[width=0.7\linewidth]{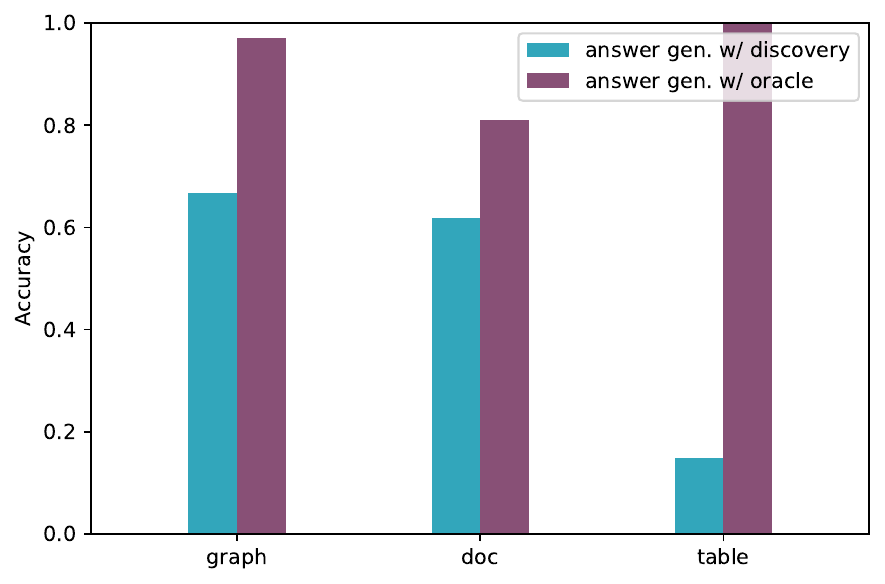}
    \caption{Task execution performance degrades due to poor data discovery despite the purported proficiency of LLMs. }
    \label{fig:answer_gen}
    \vspace{-10pt}
\end{figure}

\subsubsection{Task Execution Performance.}
We now explore the impact of data discovery performance on the downstream task execution accuracy. Figure~\ref{fig:answer_gen} captures the task execution accuracy for two scenarios: oracle and discovery. Oracle is the case where the ground truth discoverable element is provided to the LLM (\eg GPT-3.5 turbo) to provide the final answer to a question. Discovery captures the scenario where elements retrieved by the best-performing document, sub-graph, and table discovery models are provided to the LLM. We observe a significant drop in performance from \emph{oracle} to the \emph{discovery} scenario,  showcasing a $46\%$ decrease in accuracy.

%% file: tables/source_selection.tex
\begin{table*}[t]
\caption{Performance of source discovery models, both overall and for individual modalities.}
\vspace{-10pt}
\scalebox{0.9}{
\begin{tabular}{lcccccccccccc}
\toprule
& \multicolumn{3}{c}{Overall (macro average)} & \multicolumn{3}{c}{Doc} & \multicolumn{3}{c}{Table} & \multicolumn{3}{c}{graph} \\
\cmidrule(r){2-4} \cmidrule(r){5-7} \cmidrule(r){8-10} \cmidrule(r){11-13}
& P & R & F1 & P & R & F1 & P & R & F1 & P & R & F1 \\
\midrule
Select-doc & - & - & - & 66.67 & 100.00 & 80.00 & - & - & - & - & - & - \\
Select-table & - & - & - & - & - & - & 28.57 & 100.00 & 44.44 & - & - & - \\
Select-graph & - & - & - & - & - & - & - & - & - & 42.86 & 100.00 & 60.00 \\
Select-random & 45.82 & 32.80 & 36.79 & 70.73 & 34.52 & 46.40 & 28.26 & 36.11 & 31.71 & 38.46 & 27.78 & 32.26 \\
Select-all & 46.03 & 100.00 & 61.48 & 66.67 & 100.00 & 80.00 & 28.57 & 100.00 & 44.44 & 42.86 & 100.00 & 60.00 \\
GPT-3.5-turbo + $sum_{b}$ & 62.71 & 79.59 & 67.23 & 85.25 & 61.90 & 71.72 & 50.00 & 91.67 & 64.71 & 52.87 & 85.19 & 65.25 \\
GPT-3.5-turbo + $sum_{e}$  & 61.58 & 94.18 & 73.29 & 86.17 & 96.43 & 91.01 & 50.00 & 91.67 & 64.71 & 48.57 & 94.44 & 64.15 \\
GPT-4-turbo + $sum_{b}$ & 63.45 & 79.01 & 67.08 & 85.96 & 58.33 & 69.50 & 51.56 & 91.67 & 66.00 & 52.81 & 87.04 & 65.73 \\
GPT-4-turbo + $sum_{e}$ & 62.36 & 92.86 & 73.61 & 85.11 & 95.24 & 89.89 & 52.46 & 88.89 & 65.98 & 49.51 & 94.44 & 64.97 \\
\bottomrule
\end{tabular}
} 
\label{tab:source_selection}
\end{table*}

%% file: tables/table_discovery.tex
\begin{table}[!htb]
\centering
    \vspace{-5pt}
\caption{Table discovery Performance of embedding models.}
    \vspace{-10pt}
\scalebox{0.9}{
\begin{tabular}{lcccc}
\toprule
 & precision@1 & recall@1 & recall@3 & recall@5 \\
\midrule
Ada-002 & 73.10 & 73.10 & 86.57 & 94.03 \\
BGE-base & 61.19 & 61.19 & 67.16 & 70.15 \\
BM25 & 80.60 & 80.60 & 89.55 & 89.55 \\
\bottomrule
\end{tabular}
}
\label{tab:table_emb}
\vspace{-10pt}
\end{table}

%% file: tables/nl2cypher.tex
\begin{table}[!htb]
\centering
\caption{Fine-grained discovery performance on graphs.}
    \vspace{-5pt}
\scalebox{0.9}{
\begin{tabular}{lcccccc}
\toprule
 & \multicolumn{3}{c}{node} & \multicolumn{3}{c}{edge} \\
\cmidrule(r){2-4} \cmidrule(r){5-7}
 & P & R & F1 & P & R & F1 \\
\midrule
Llama2-7b-chat & 18.18 & 19.19 & 18.67 & 0.00 & 0.00 & 0.00 \\
Llama2-70b-chat & 4.06 & 5.33 & 4.61 & 0.00 & 0.00 & 0.00 \\
GPT-3.5-turbo & 58.83 & 68.69 & 63.38 & 37.26 & 39.39 & 38.30 \\
GPT-4-turbo & 74.24 & 75.76 & 74.99 & 37.37 & 37.88 & 37.62 \\
\bottomrule
\end{tabular}}
\label{tab:fine_nl2cypher}
\vspace{-5pt}
\end{table}

%% file: tables/doc_para_discovery.tex
\begin{table*}[t]
\caption{Performance of embedding models on document and paragraph discovery task.} 
\scalebox{0.9}{
\begin{tabular}{lcccccccc}
\toprule
& \multicolumn{4}{c}{Document} & \multicolumn{4}{c}{Paragraph} \\
\cmidrule(r){2-5} \cmidrule(r){6-9}
& R-precision & recall@1 & recall@3 & recall@5 & R-precision & recall@1 & recall@3 & recall@5 \\
\midrule
BM25 & 21.43 & 15.87 & 26.19 & 28.57 & 45.50 & 29.89 & 58.99 & 71.43 \\
Ada-002 & 48.41 & 45.24 & 57.94 & 58.73 & 39.15 & 33.07 & 51.85 & 65.34 \\
BGE-base & 52.38 & 47.62 & 68.25 & 71.43 & 57.41 & 42.59 & 78.04 & 84.13 \\
BGE-large & 61.90 & 54.76 & 70.63 & 78.57 & 55.03 & 44.18 & 77.25 & 86.51 \\
\bottomrule
\end{tabular}
}
\label{tab:doc_para}
\vspace{-10pt}
\end{table*}

%% file: 5_related.tex
\section{Related Work}
\label{sec:related}

\stitle{Organizations and Data Silos.}
As reported by American Management Association~\cite{comfort2005risk}, data silos in modern organizations are unavoidable and have a negative impact on aspects such as business decision-making. 
With the proliferation of e-commerce-driven applications, data in organizations sit in hundreds of applications or servers that result in data silos~\cite{patel2019bridging}. Data silos are formed due to team dynamics, organizational structure, and data culture~\cite{vertesi2011value}. 
As mentioned in Section~\ref{sec:overview}, we observed similar data cultures within large organizations in the HR domain, where application-specific silos emerge over time. We take these observations into account while modeling the data sources in the benchmark.

\stitle{Data Discovery and Modalities.} Existing benchmarks are typically designed to evaluate unimodal data discovery models --- operating over unimodal datalakes --- such as tables~\cite{srinivas2023lakebench} and documents~\cite{Petroni2020KILTAB} on a variety of tasks. For example, unionable and joinable search for tables and question answering and fact-checking for documents. While recent work explores a cross-modal data discovery system, CMDL~\cite{eltabakh2023cross}, they only consider two modalities within the data lake --- tables (\eg pharmaceutical databases) and textual data (\eg PubMed paper abstracts.) VerifAI~\cite{tang2023verifai} fact-checks and verifies LLM generated text by leveraging information from a data lake containing Wikipedia documents and tables. CMDL~\cite{eltabakh2023cross} exploits a pre-computed mapping across sources to compute a joint embedding for retrieval and VerifAI~\cite{tang2023verifai} employs ElasticSearch~\cite{tong2015elasticsearch} to index tables or
text files serialized as strings. Unlike the aforementioned data lakes, we explore an additional modality, \ie graphs, and new discovery elements both coarse-grained (\eg \emph{source modality} and fine-grained (\eg node, path, sub-graph.) 

\stitle{Compound AI System Evaluation.} Evaluation frameworks for compound AI systems such as retrieval-augmented-generation (RAG) applications such as ARES~\cite{saad2023ares} and RAGAS~\cite{es2023ragas} are reliant on synthetic data generation and are limited to documents only. Moreover, these frameworks prioritize measuring downstream task performance such as answer relevance and context relevance. While RAGAS~\cite{es2023ragas} measures document retrieval, aspects such as source discovery and additional modalities, \eg tables and graphs, are not considered as part of the evaluation objective. 

%% file: 6_conclusion.tex
\section{Discussion and Conclusion}
\label{sec:conclusion}

The database community has always adjusted to changes in technology often demonstrated through the release of benchmarking standards --- from pioneering the data warehouse-era TPC-H to modern decision support system-focused benchmarks such as TPC-DS~\cite{nambiar2006making}. With the emergence of LLM-powered agentic workflows in compound AI systems, the industry is again embarking on a new approach within data platforms. Within such a compound AI system, the range of possible execution strategies is vast. For example, even in the discovery 
scenarios explored in Section~\ref{sec:exp}, there are: (i) many discovery methods and language models to choose from and (ii) the efficiency of discovery methods varies with data source type and task difficulty. Therefore,  effective navigation of the vast space to find a good design often hinges on these benchmarks. 

In this work, we explored a CAS, LlamaIndex, in a constrained setting. The immediate follow-up work can focus on scaling up the data sources and benchmarks. As the scale of the data increases, automatic or semi-automatic approaches to data preparation and benchmark creation are required. For example, new data can be incrementally added in the future to include other domains (\eg NFL and MLS) using the same semi-automatic data collection strategy outlined in the paper. Integrating additional collections in an existing data source (\,  e.g., basketball history and statistics~\cite{Basketball}) remains a future work. Morover, new trade-offs involving storage, latency, and accuracy may emerge in an attempt to maintain the effectiveness of the discovery methods over large-scale data. 

\stitle{Concluding Remarks.} In this work, we propose \system, a benchmark for evaluating the performance of data discovery agents in compound AI systems operating on siloed enterprise data. Our experiments conducted on carefully designed benchmarks and data unveil the vast design space of agentic workflows, even in the context of data discovery tasks. 